\documentstyle[12pt]{article}
\newcommand{\be}{\begin{equation}}
\newcommand{\ee}{\end{equation}}
\newcommand{\bn}{\begin{eqnarray}}
\newcommand{\en}{\end{eqnarray}}
\newcommand{\bal}{\mbox{\boldmath$\alpha$}}

\begin{document}
\begin{titlepage}
\vspace{1cm}
\begin{center}
\vspace*{1.0cm}

{\Large  Polynomial Algebras and Higher Spins}

\vskip 2.0cm

by

\vskip 0.5 cm

{\large {\bf M. Chaichian}}

\vskip 0.5cm

High Energy Physics Laboratory, Department of Physics \\
and Research Institute for High Energy Physics, \\
University of Helsinki\\
P.O.Box 9 (Siltavuorenpenger 20 C), FIN-00014,  \\
Helsinki, Finland

\vskip 0.2cm

$and$

\vskip 0.2cm

{\large {\bf A.P.Demichev}}\renewcommand{\thefootnote}
{\dagger}\footnote{on leave of absence from
Nuclear Physics Institute,
Moscow State University,
119899, Moscow, Russia}

\vskip 0.5cm

Centro Brasileiro de Pesquisas Fisicas - CBPF/CNPq, \\
Rua Dr.Xavier Sigaund, 150, \\
22290-180, Rio de Janeiro, \\
RJ Brasil

\end{center}

\vspace{3 cm}

\begin{abstract}
\normalsize

Polynomial relations for generators of $su(2)$ Lie algebra in arbitrary
representations are found. They generalize usual relation for Pauli
operators in spin 1/2 case and permit to construct modified
Holstein-Primakoff transformations in finite dimensional Fock spaces. The
connection between $su(2)$ Lie algebra and q-oscillators with a root of
unity q-parameter is considered. The meaning of the polynomial relations
from the point of view of quantum mechanics on a sphere are discussed. 

\end{abstract}
\end{titlepage}

\section{Introduction}

Quantum mechanics of spin systems plays outstanding role both in
applications and developments of new theoretical constructions and
methods. The standard way of quantization of a spin systems is to consider
spin components $\{{\bf J}_i\}_{i=1}^3$ as generators of $su(2)$ Lie
algebra and to use its well known representation theory. Theoretically
more profound method \cite{Ber} describes spins as quantum mechanical
systems with curved phase spaces which are orbits (spheres) of a group
$SU(2)$ in a space dual to the $su(2)$ Lie algebra space. This
construction, in particular, permits to obtain quasi-classical
approximation for spin systems in the limit of infinite spin value $j$,
the inverse spin value $1/j$ playing the role similar to that of the
Planck constant $\hbar$ for a Heisenberg-Weyl algebra. The fact that the
phase space of a spin system is a sphere leads to the constraint
\be
\sum_1^3 {\bf J}^2_i = j(j+1)\ ,                      \label{1}
\ee
for any given representation labeled by a spin value $j$. The presence of
the constraint causes computational difficulties in some cases and one of
ways to resolve it, well known in solid bodies theory \cite{Kit}, is to
express the spin operators ${\bf J}_i$ in terms of bosonic creation and
annihilation operators ${\bf b^+,b}$ with help of Holstein-Primakoff (HP)
transformation \cite{HP}
\bn
{\bf J}_+ =\sqrt{2j}\left( 1-\frac{{\bf N}}{2j}\right) ^{1/2} {\bf b} \ ,
\nonumber \\
{\bf J}_- =\sqrt{2j}{\bf b}^+\left( 1-\frac{{\bf N}}{2j}\right) ^{1/2} \ ,
\label{2} \\
{\bf J}_3 = j-{\bf N} \ ,\qquad {\bf N} = {\bf b^+b} \ ,\nonumber
\en
$$
[{\bf b},{\bf b}^+ ] = 1 \ .
$$

This transformation considerably simplifies operator calculations but has
two obvious shortcomings: ({\it i}) it contains square roots and forces to
use series expansions for Hamiltonians and any other expressions; ({\it
ii}) though representation spaces for spin operators ($su(2)$ algebra) has
finite dimensions $d=2j+1$, the Fock space for ${\bf b^+,b}$ operators is
infinite dimensional one. Both these reasons show that HP transformation
(\ref{2}) is correct when mean values $N$ of excitation number operator
${\bf N}$ satisfy the condition $N\ll j$. 

From the other hand, for spin $1/2$ case there is exact relation 
between spin
operators ${\bf J}_i$ and Pauli matrices $\sigma^\pm$ which satisfy
anti\-commuta\-tion relation
\be
\sigma^+\sigma^- + \sigma^-\sigma^+  = 1 \ .         \label{3}
\ee
Taking into account (\ref{3})  one can put for $j=1/2$
\be
{\bf J}_\pm = \sigma^\pm \ , \qquad {\bf J}_3 = {\bf N}-\frac{1}{2}
,\qquad {\bf N}=\sigma^+\sigma^- \  .
                                  \label{4}
\ee

So one can use the elements $\sigma^\pm$ with CR (\ref{3}) as spin $1/2$
dynamical variables without any constraints. Moreover, these variables
permit to consider the system as the result of the quantization of a
"classical" mechanics with anticommuting variables \cite{BerM} and to
construct path integral representation for the evolution operator of the
system with help of integral over Grassmann anticommuting variables (see
e.g. \cite{Ber2}). 

In this sense the $j=1/2$ representation seems to be distinguished from
any other and it is desirable to find appropriate generalization of the
relations (\ref{3}), (\ref{4}) for any $j$. 

It is well known that q-oscillators \cite{qOsc} for $q^K=-1$ ($K$ is
integer number) have finite dimensional Fock space, the dimension being
dependent on the value of the parameter $q$. So it is natural to try to
express the generators ${\bf J}_i$ in arbitrary but definite
representation in terms of q-oscillator operators ${\bf a^+,a,N}$ with
defining relation
\be
{\bf aa^+}- q{\bf a^+a} = q^{- {\bf N}}                    \label{5}
\ee
\be
[ {\bf N,a^+}]= {\bf a^+}\ , \qquad q^K = -1\ .       \label{6}
\ee
Note that because of the exponent in the RHS of (\ref{5}) the algebra of
${\bf a^+,a}$ operators is non-quadratic. But if q-parameter is equal to
root of unity, the relation (\ref{5}) is polynomial due to nilpotence of
${\bf a^+,a}$ (finite dimensionality of the Fock space). One can check
easily that for $q^2 = -1$ the relation (\ref{5}) is equivalent to
(\ref{3}) and hence for $j=1/2$ spin operators are expressed in terms of
q-oscillator with $q^2=-1$.

 Fock space representation of (\ref{5}),(\ref{6}) has a form \cite{qOsc}
 \bn
 {\bf a}|n\rangle=\sqrt{[n]_q}|n-1\rangle \ ,      \nonumber \\
 {\bf a^+}|n\rangle=\sqrt{[n+1]_q}|n+1\rangle \ ,      \label{7} \\
{\bf N}|n\rangle=n|n\rangle \ ,      \nonumber
\en
$$
[n]_q=\frac{q^n-q^{-n}}{q-q^{-1}}
$$
Finite dimensionality of the Fock space follows from the fact that $[K]_q=0$
if $q^K=-1$.

In \cite{ChaD} we constructed path integral representations for the
evolution operator of the q-oscillators. Thus the expression for spin
components in terms of q-oscillator operators implies the construction of
path integrals for higher spins dynamics. 

The aim of this letter is to present the relation between spin operators
for arbitrary $j$, q-oscillators and other non-quadratic but polynomial
algebras. In Conclusion we discuss the meaning of the obtained relations
from the point of view of quantum mechanics on a sphere.

\section{Algebra $su(2)$ and q-oscillators}

As we mentioned already, the CR for q-oscillator with $q^2=-1$ are equal
to the relations for the Pauli matrices. Let us consider the next value of
integer parameter $K$ for the q-oscillators, i.e. $K=3$, and, hence,
$q^3=-1$. In this case the Fock space is 3-dimensional and operator
$q^{-{\bf N}}$ can be written in the form

\be
q^{-{\bf N}}= C_0 + C_1{\bf a^+a} + C_2{\bf a^{+^2}a^2}\ .    \label{8}
\ee

Comparing the action of LHS and RHS of (\ref{8}) on arbitrary vector of
Fock space one finds
\be
q^{-{\bf N}}= 1 - q{\bf a^+a} - {\bf a^{+^2}a^2}\ .    \label{9}
\ee
Substitution of this relation in (\ref{5}) gives
\be
{\bf aa^+} + {\bf a^{+^2}a^2} = 1\ .                    \label{9a}
\ee
This seems to be very natural generalization of relation (\ref{3}) for
Pauli matrices.

Representation (\ref{7}) of ${\bf a^+,a}$ in this case have very
simple form
\be
{\bf a^+}=
\left(
\begin{array}{ccc}
0 & 1 & 0 \\
0 & 0 & 1 \\
0 & 0 & 0
\end{array}
\right) \ , \qquad
 {\bf a}=
\left(
\begin{array}{ccc}
0 & 0 & 0 \\
1 & 0 & 0 \\
0 & 1 & 0
\end{array}
\right) \ ,                                            \label{11}
\ee
We immediately recognize in (\ref{11}) generators ${\bf J}_\pm$ in spin 1
representation up to trivial rescaling. Now one can find ${\bf N}$ in the
same way like $q^{-{\bf N}}$
$$
{\bf N} ={\bf a^+a} + {\bf a^{+^2}a^2} \ ,
$$
so that
$$
[{\bf a,a^+}] = 1 - {\bf N} \ ,
$$
and identify
\be
{\bf J}^{(1)}_+ = {\bf a^+}/\sqrt{2} \ , \quad
{\bf J}^{(1)}_- = {\bf a}/\sqrt{2}  \ , 
\quad {\bf J}^{(1)}_3 = ({\bf N} - 1)/2
\ ,                                                      \label{11aa}
\ee
where superscript index indicates the representation.

As ${\bf J}_\pm$ coincide with ${\bf a^+,a}$ for $K=2,3$ one can assume
that analogous relations hold for all values of $K$. Unfortunately, this
is not the case. Indeed, considering just the next value $K=4$ one finds
$$
{\bf aa^+} = 1+(\sqrt{2} - 1){\bf a^+a}-(\sqrt{2} - 1){\bf a^{+^2}a^2}+
(\sqrt{2} - 2){\bf a^{^3+}a^3} \ .
$$
But
$$
[{\bf a,a^+}]\neq A+B{\bf N}
$$
for any constants $A,B$.

Thus to establish the CR  for ${\bf J}^{(n)}_\pm ,\ n>1$ one
must consider more general polynomial algebras than q-oscillators.

\section{Polynomial algebras with finite dimensional Fock
space representations}

Consider general condition of p-dimensional Fock space representation
for creation
$\bal^+$ and destruction $\bal$ operators
\be
\bal^{+^{\bf p}}|0\rangle = 0                           \label{11a}
\ee
and write $\bal\bal^+$ in normal form that defines the CR
\be
\bal\bal^+ = \sum^{p-1}_{l=0}C_l\bal^{+^{\bf l}}\bal^{\bf l}     
\label{12}
\ee
with the  undefined so far constants $C_l$. Acting on identity (\ref{11a})
by $\bal$ from the left one obtains condition for $C_l$. For example,
in the case $p=4$ the condition is (without restriction of generality one
can always put $C_0=1$ by rescaling of $\bal,\bal^+$)
\bn
C_1^3C_2+C_1^3C_3+C_1+C_1^2C_2^2+C_1^2C_2C_3 + 
3C_1^2C_2+2C_1^2C_3+C_1^2 +
                                                            \nonumber\\
+2C_1C_2^2+2C_1C_2C_3 +
3C_1C_2+2C_1C_3+C_1+C_2^2+C_2C_3+C_2+C_3+1=0
                                                              \label{13}
\en
Thus we have two parametric family of CR (\ref{12}). The obvious solution
of (\ref{13}) is
$$
C_1=C_2=0 \ , \qquad C_3=-1 \ ,
$$
which corresponds to CR
\be
\bal\bal^++\bal^{+^{\bf 3}}\bal^{\bf 3}=1 \ .              \label{14}
\ee
This CR is similar to q-oscillator CR for the cases $K=2,3$. The
representation of the CR (\ref{14}) is
\be
\bal^+=
\left(
\begin{array}{cccc}
0 & 1 & 0 & 0 \\
0 & 0 & 1 & 0 \\
0 & 0 & 0 & 1 \\
0 & 0 & 0 & 0 \\
\end{array}
\right) \ , \qquad
 \bal =
\left(
\begin{array}{cccc}
0 & 0 & 0 & 0 \\
1 & 0 & 0 & 0 \\
0 & 1 & 0 & 0 \\
0 & 0 & 1 & 0 \\
\end{array}
\right) \ ,                                            \label{15}
\ee
It looks like natural generalization of Pauli matrices $\sigma^\pm$. 
It is tempting to suppose now that all algebras with CR
\be
\bal\bal^++\bal^{+^{\bf p}}\bal^{\bf p}=1 \                \label{16}
\ee
have representations of the dimensions $p+1$. 
This is true because it is easy to check
that (\ref{16}) is consistent with (\ref{11a}) indeed. The representation
of (\ref{16}) is also straightforward generalization of the Pauli matrices
$\sigma^\pm$, representations (\ref{11}) and (\ref{15}) for $p=2$ and $p=3$
cases
\be
(\bal^+)_{ij} = \delta_{i,j-1} \ ,\quad (\bal)_{ij} = \delta_{i-1,j}\ , \quad
i,j=1,...,p+1\ .                                        \label{17}
\ee
Unfortunately, the simplest possible form of creation and destruction
operators (\ref{17}) with CR (\ref{16}) do not coincide with ${\bf
J}^{(n)}_\pm$ for $n>1$. Excitations number operator ${\bf N}$ has the
form
\be
{\bf N} = \sum^p_{l=1}\bal^{+^{\bf l}}\bal^{\bf l}               \label{18}
\ee
and as in the case of q-oscillators for $K>3$
$$
[\bal,\bal^+]\neq A+B{\bf N}
$$
for any $A,B$.

There are two ways to overcome the problem. We can use the algebra
(\ref{16}) and construct polynomial relations between ${\bf J}_\pm$ and
$\bal^+,\bal$. This is analogous to the Holstein-Primakoff transformation
but with two crucial distinctions which remove the shortcomings of the
original HP transformation: both sets of operators act now in the same
Fock space and the transformation we shall construct does not contain
square roots. So {\it a priori} there are no needs in any approximations. 

We illustrate the construction on the example of $p=4$. The generalization
to the higher values of $p$ is straightforward. 

Well known representation of generators ${\bf J}_\pm$ have a form
\bn
{\bf J}_+|n\rangle = \gamma_{n+1}|n+1\rangle \ ,
{\bf J}_-|n\rangle = \gamma_n|n-1\rangle \ ,       \label{19} \\
\gamma_{n+1}=\sqrt{(2j-n)(n+1)/2} \ .      \nonumber
\en
Due to finite dimensionality of the Fock space (nilpotence of
$\bal,\bal^+$) one can write
\be
{\bf J}_+ = \bal^+(D_0+D_1\bal^+\bal+D_2\bal^{+^{\bf 2}}\bal^{\bf 2}) \ .
\label{20}
\ee
Acting by ${\bf J}_+$ in the form (\ref{20}) on Fock space vectors and
comparing with
(\ref{19}) one obtains

\be
{\bf J}_+ =\bal^+\left[\gamma_1+(\gamma_2-\gamma_1)\bal^+\bal+(\gamma_3-
\gamma_2)\bal^{+^{\bf 2}}\bal^{\bf 2}\right]  \ .
  \label{21}
\ee

Generator ${\bf J}_-$ can be found by hermitian conjugation
${\bf J}_-=({\bf J}_+)^\dagger$ and generator ${\bf J}_3$ by
commutation of ${\bf J}_+$ and ${\bf J}_-$
$$
{\bf J_3}=\frac{1}{2}[{\bf J_+,J_-}] = {\bf N}-\frac{3}{2} \ ,
$$
where ${\bf N}$ is defined by (\ref{18}).

Another way to express ${\bf J}_\pm$ generators through creation and
destruction operators is to consider more general solution of (\ref{13})
together with conditions
\be
[{\bf a,a^+}] = A + B{\bf N} \ ,                     \label{22}
\ee
\be
[{\bf N,a^+}] = {\bf A^+} \ ,                     \label{23}
\ee
where
$$
{\bf N} = F_1{\bf a^+a}+F_2{\bf a^{+^2}a^2}+F_3{\bf a^{+^3}a^3} \ .
$$
After straightforward but lengthy calculations one finds 
that the equations (\ref{13}),(\ref{22}) and (\ref{23}) have the unique 
solution \bn
{\bf aa^+} =1+ \frac{1}{3}{\bf a^+a}-\frac{1}{3}{\bf a^{+^2}a^2}
-\frac{2}{3}{\bf a^{+^3}a^3} \ ,                    \label{24} \\
{\bf N} = {\bf a^+a}+\frac{1}{2}{\bf a^{+^2}a^2}+{\bf a^{+^3}a^3} \ ,
\nonumber \\
A=1 \qquad B=-\frac{2}{3} \nonumber
\en
Trivial rescaling ${\bf J}_+=\sqrt{3}{\bf a^+},\
{\bf J}_-=\sqrt{3}{\bf a}$ and definition ${\bf J}_3 = {\bf N}-\frac{3}{2}$
turns (\ref{22}), (\ref{23}) into the CR of Lie algebra $su(2)$. The algebras
for higher spins representations can be found analogously.

\section{Conclusion}

We studied the closed algebras of $su(2)$ generators ${\bf J}_\pm$ in
arbitrary representations, suggested two methods to find them and
presented explicit formulas for $j=1$ and $j=3/2$ cases. The algebras of
${\bf J}^{(1)}_\pm$ for $j=1$ case coincides with those for q-oscillator ,
q-parameter being a third root of minus unity $q^3=-1$. From the other
hand this algebra looks like natural generalization of the algebra of
$\sigma^\pm$ Pauli matrices. The algebras of ${\bf J}_\pm$ for $j\neq
1/2,1$ do not coincide with standard q-oscillators. Instead, we suggested
natural generalization of the CR in the form (\ref{16})
$$
\bal\bal^++\bal^{+^{\bf p}}\bal^{\bf p} = 1
$$
and proved that it has representation in p-dimensional Fock space of the
very simple form (\ref{17}). There are polynomial relations between ${\bf
J}_\pm$ and $\bal$-operators which can be considered as improved
Holstein-Primakoff transformations. 

It is instructive to discuss the connection between expression for a
$su(2)$ algebra generators in terms of $\bal,\bal^+$-operators and well
known Jordan-Schwinger representation \cite{Schw}
\be
{\bf J}_+ = {\bf b_1^+b_2}\ , \quad {\bf J}_- = {\bf b^+_2b_1}\ ,
\quad {\bf J}_3 = \frac{1}{2}({\bf n_1} - {\bf n_2})/2 \ ,
{\bf n}_i = {\bf b}_i^+{\bf b}_i\ ,                  \label{25}
\ee
where ${\bf b}_i^+,{\bf b}_i\ (i=1,2)$ are two pairs of Bose creation and
destruction operators with the usual "flat" CR
\be
[{\bf b}_i,{\bf b}^+_j]=\delta_{ij} \qquad i,j=1,2\ .                \label{26}
\ee
From the point of view of quantization on a Lie algebra $su(2)$ this
representation can be understood as embedding of the three dimensional
space $su(2)$ with singular symplectic form in the four dimensional phase
space of the two oscillators. The latter can be quantized in the usual way
(\ref{26}). Note that the general method of quantization of systems with
curved phase space via embedding in higher dimensional flat phase space
was considered in \cite{BatF}. In a sense this way is opposite to that
considered in \cite{Ber} where the classical mechanics on $su(2)$ is
restricted to two dimensional sphere and then quantized. 

The restriction of the algebra of all operators generated by ${\bf
b}_i,{\bf b}^+_i$ to those generated by ${\bf J_{\pm},J_3}$ according to
(\ref{25}) leads to the two main results from the point of view of
representations in Hilbert spaces: ({\it i}) the whole Fock space for the
two oscillators becomes reducible for the subalgebra; ({\it ii}) some
operators of the subalgebra become proportional to unity in irreducible
subspaces. The first fact leads in turn to finite dimensional Hilbert
spaces of the spin systems and, hence, to the absence of a classical limit
for them. The second fact is well known for the Casimir operator ${\bf
C}=\sum_{i=1}^3 {\bf J}_i^2$. It is central element of the subalgebra and
so proportional to unity in {\it any} irreducible representation. However,
as we have shown in this paper, for any given representation there is
another operator constructed from ${\bf J}_{\pm}$ which is equal to unity.
For example, for the two dimensional representation this is the operator
${\bf J_-J_+}+{\bf J_+J_-}$, for the three dimensional one it has the form
$(1/2){\bf J_-J_+}+(1/4){\bf J_+^2J_-^2}$ (according to (\ref{9a}),
(\ref{11aa})). For higher dimensional representations these operators can
be obtained with help of relation (\ref{16}) for the
$\bal^+,\bal$-operators (one can easily express $\bal^+,\bal$ in terms of
${\bf J}_\pm$ due to nilpotence). The crucial distinction from the Casimir
operator is that these identity operators are different for different
irreducible representations. Now we can further restrict the algebra to
those generated by ${\bf J}_\pm$ or $\bal^+,\bal$ operators only, with
defining relation of the type (\ref{16}). This can be interpreted as the
reduction on a spherical phase space {\it after} quantization (as opposite
to the scheme in \cite{Ber} where the reduction is made {\it before}
quantization). 

As a final remark, let us note that quantum mechanics for spin 1/2 can be
considered formally as a quantization (central extension) of the
"classical" Grassmann algebra
$$
z\bar z +\bar z z=0, \qquad z^2=\bar z^2=0 \ .
$$
In the same way, the quantum mechanics for higher spins can be considered
as the quantization of the algebras with the following defining relations
$$
z\bar z + \bar z^p z^p = 0, \qquad z^{p+1}=\bar z^{p+1}=0 \ ,
$$
which follows from (\ref{16}) and nilpotence of the operators $\bal^+,\bal$.
\vskip 10 mm
\centerline{{\bf Acknowledgements}}
We thank P. Kulish for useful discussions.
A.D.'s work was partially supported by the INTAS-93-1630 grant.

\end{document}